\documentclass{svproc}
\usepackage{url}
\usepackage{graphicx}
\usepackage{enumitem}
\usepackage{adjustbox}

\begin{document}
\mainmatter              
\title{Modern CNNs for IoT Based Farms}
\titlerunning{Modern CNNs for IoT Based Farms}  
%
\author{Patrick Kinyua Gikunda\inst{1,2} \and
Nicolas Jouandeau\inst{2}}
\authorrunning{Gikunda and Jouandeau.} 

\institute{
Department of Information Technology \\ Dedan Kimathi University of Technology, Kenya \\
\email{patrick.gikunda@dkut.ac.ke} \and
Le Laboratoire d'Informatique Avancée de Saint-Denis (LIASD) \\ University of Paris 8, France\\
\email{n@ai.univ-paris8.fr}
}

\maketitle              

\begin{abstract}
Recent introduction of ICT in agriculture has brought a number of changes in the way farming is done. This means use of Internet of Things(IoT), Cloud Computing(CC), Big Data (BD) and automation to gain better control over the process of farming. As the use of these technologies in farms has grown exponentially with massive data production, there is need to develop and use state-of-the-art tools in order to gain more insight from the data within reasonable time. In this paper, we present an initial understanding of Convolutional Neural Network (CNN), the recent architectures of state-of-the-art CNN and their underlying complexities. Then we propose a classification taxonomy tailored for agricultural application of CNN. Finally, we present a comprehensive review of research dedicated to applications of state-of-the-art CNNs in agricultural production systems. Our contribution is in two-fold. First, for end users of agricultural deep learning tools, our benchmarking finding can serve as a guide to selecting appropriate architecture to use. Second, for agricultural software developers of deep learning tools, our in-depth analysis explains the state-of-the-art CNN complexities and points out possible future directions to further optimize the running performance.
\end{abstract}
\section{Introduction} \label{sec:intro}

The global population is set to touch 9.6 billion mark by year 2050 \cite{Godfray_2010}. The continous population growth means increase in demand for food to feed the population \cite{Lutz_2008}. Agriculture is the practice of cultivation of land and breeding of animals \& plants to provide food and other products in order to sustain and enhance life \cite{Bruinsma_2003}. Due to the  extreme weather conditions, rising climate change and environmental impact resulting from intensive farming practices \cite{Gebbers_2010}, farmers are now forced to change their farming practices. To cope with the new farming challenges, farmers are forced to practice smart farming \cite{Krintz_2016}, which offers solutions of farming management and environment management for better production. Smart farming focuses on the use of information and communication technology(ICT) in the cyber-physical farm management cycle for efficient farming \cite{Kamilaris_2018}.

Curent ICT technologies relevant for use in smart farming include IoT \cite{Weber_2010}, remote sensing \cite{Anindya_2016}, CC \cite{Jinbo_2018} and BD \cite{Chi_2016}. Remote sensing is the science of gathering information about objects or areas from a distance without having physical contact with objects or areas being investigated. Data collected through remote sensing and distributed devices is managed by cloud computing technology, which offers the tools for pre-processing and modelling of huge amounts of data coming from various heterogeneous sources \cite{Waga_2017}. These four technologies could create applications to provide solutions to todays's agricultural challenges. The solutions include real time analytics required to carry out agile actions especially in case of suddenly changed operational or environmental condition (e.g. weather or disease alert). The continous monitoring, measuring, storing and analysing of various physical aspects has led to a phenomena of big data \cite{Chen_2014}. To get insight for practical action from this large type of data requires tools and methods that can process multidimensional data from different sources while leveraging on the processing time. 

One of the sucessful data processing tool applied in this kind of large dataset is the biologically inspired Convolutional Neural Networks (CNNs),  which have achieved state-of-the-art results \cite{Russakovsky_2015} in computer vision \cite{Ming_2015} and data mining \cite{Poria_2016}. As deep learning has been successfully applied in various domains, it has recently entered in the domain of agriculture \cite{Kamilaris_2018}. CNN is a subset method of Deep Learning(DL) \cite{Goodfellow_2015}, defined as deep, feed-forward Artificial Neural Network(ANN)\cite{Bhandare_2016}. The CNN covolutions allow data representations in a hierarchical way \cite{Schmidhuber_2015}. The common characteristics of CNN models is that they follow the same general design principles of successive applying convolutional layers to the input, periodically downsampling the spatial dimensions while increasing the number of feature maps. These architectures serve as rich feature extractors which can be used for image classification, object detection, image segmentation and many more other advanced tasks. This study investigates the agricultural problems that employ the major state-of-the-art CNN archtectures that have participated in the ImageNet Large Scale Visual Recognition Challenge (ILSVRC) \cite{LSVRC_2018} with highest accuracy in a multi-class classification problem. ImageNet \cite{ImageNet_2018} classification challange has played a critical role in advancing the CNN state-of-the-art \cite{Russakovsky_2015}. The motivation for carrying out the study include: a) CNNs has better precision compared to other popular image-processing techniques in the large majority of problems \cite{Zahangir_2018}. b) CNN has entered in the agricultural domain with promising potential \cite{Kamilaris_2018_1}. c)  all the CNN models that have achieved the top-5 error are successful when applied in other computer vision domain with remarkable results \cite{Zahangir_2018}. This review aims to provide insight on use of state-of-the-art CNN models in relation to smart farming and to identify smart farming research and development challenges related to computer vision. Therefore the analysis will primarily focus on the success on use of state-of-the-art CNN models in smart farms with a intention to provide this relevant information to the future researchers. From that perspective the research questions to be addressed in the study are:

\begin{enumerate}[label=(\alph*)]
  \item What is the role of CNN in smart farming?
  \item What type of the state-of-the-art CNN architecture should be used?
  \item What are the benefits of using state-of-the-art CNN in IoT based agricultural systems?
\end{enumerate}

The rest of the paper is organized as follows: in Section 2, we present an overview of existing  state-of-the-art CNNs  architectures including their recent updates. Then we propose a taxonomy to provide a systematic classification of agricultural issues for CNN application. In Section 3, we present the existing state-of-the-art CNNs and their scope of application in agri-systems. We conclude the paper in Section 4.

\section{Methodology} \label{sec:Methodology}

In order to address the research questions a bibliographic analysis in the domain under study was done between 2012 and 2018, it involved two steps: a) collection of related works and, b) detailed review and analysis of the works. The choice of the period is from the fact that CNN is rather a recent phenomenon. In the first step, a keyword-based search  using all combinations of two groups of keywords of which the first group addresses CNN models (LeNet, AlexNet, NIN, ENet, GoogLeNet, ResNet, DenseNet, VGG, Inception) and the second group refers to farming (i.e. agriculture, farming, smart farming). The analysis was done while considering the following research questions: a) smart farm problem they addressed, b)dataset used, c) accuracy based on author's performance metric, d) state-of-the-art CNN model used. Its important to note that use of state-of-the-art deep learning has great potential, and there have been recent small comparative studies to analyse and compare the most efficient archtecture to use in agricultural systems. They include: Comparison between  LeNet, AlexNet, and VGGNet on automatic identification of center pivot irrigation \cite{Zhang_2018} and comparison between VGG-16, Inception-V4, Resnet and DenseNet for plant disease identification \cite{Chebet_2018}.

\section{State-of-the-art CNN: An Overview} \label{sec:cnn}
\begin{figure}[h!]
 \centering
  \includegraphics[height=8cm,height=4.5cm]{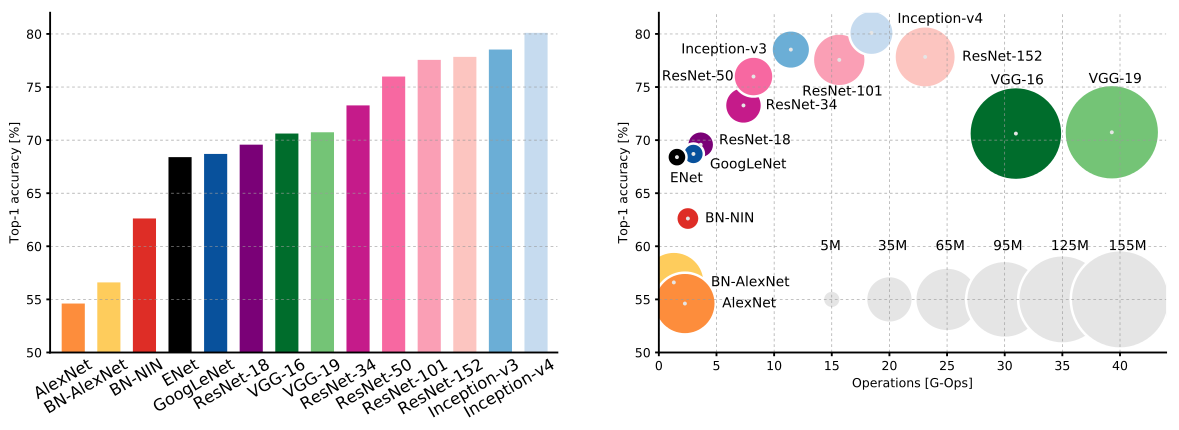}
 \caption{\small{Top-1 Accuracy vs the computational cost. The size of the circles is proportaional to number of parameters. Legend;the grey cirles at the botton right represents number of parameters in millions. \cite{Canziani_2016}}}
 	\label{fig:cr}
 \end{figure}

CNNs typically perform best when they are large, meaning when they have more deeper and highly interconnected layers \cite{Renzo_2018}. The primary drawback of these archtectures is the computational cost, thus large CNNs are typically impractically slow especially for embedded IoT devices \cite{Canziani_2016}. There are recent research efforts on how to reduce the computation cost of deep learning networks for everyday application while maintaining the prediction accuracy \cite{HasanPour_2018}. In order to understand the application of the state-of-the-art CNN archtectures in agricultural systems, we reviewed the accuracy and computational requirements from relevant literature including recent updates of networks as shown in Fig \ref{fig:cr}. The classical state-of-the-art deep network architectures include; LeNet \cite{Lecun_1998}, AlexNet \cite{Alex_2012}, NIN \cite{Lin_2013}, ENet\cite{Paszke_2016}, ZFNet \cite{Zeiler_2013}, GoogleLeNet \cite{Szegedy_2015} and VGG 16 \cite{Simonyan_2014}. Modern architectures include; Inception \cite{Szegedy_2016}, ResNet \cite{Kaiming_2015}, and DenseNet \cite{Huang_2017}. 

LeNet-5 is a 7-layer pioneer convolutional network by LeCun et al. \cite{Lecun_1998} to classify digits, used to recognise hand-written numbers digitized in 32x32 pixel greyscale input images. High resolution images require more convolutional layers, so the model is constrained by the availability of the computing resources. 

AlexNet is a 5-layer network similar to LeNet-5 but with more filters \cite{Alex_2012}. It outperformed Lenet-5 and won the LSVRC challenge by reducing the top-5 error  from 26.2\% to 15.3\%. Use Rectified Linear Unit (Relu) \cite{Xu_2015} instead of Hyperbolic Tangent (Tanh) \cite{Luo_2016} to add non-linearity and accelerates the speed by 6 times. Droupout was employed to reduce over-fitting in the fully-connected layers. Overlap pooling was used to reduce the size of the network while reducing top-1 error by 0.4\% and top-5 error by 0.3\%. 

Lin et al. \cite{Lin_2013} created a Network in Network(NIN) which inspired the inception architecture of googlenet. In their paper, they replaced the linear filters with nonlinear multi linear perceptrons that had better feature extraction and accuracy. They also replaced the fully connected layers with activation maps and global average pooling. This move helped reduce the parameters and network complexity. 

In their article Paszke et al. \cite{Paszke_2016} introduced an Efficient Neural Network (ENet) for running on low-power mobile devices while achieving state-of-the-art results. ENet architecture is largely based on ResNets. The structure has one master and several branches that separate from the master but concatenate back. 

In their work Zeiler and Fergus \cite{Zeiler_2013} created ZFNet which won a ILSVRC 2013 \cite{LSVRC_2013} image classification. It was able to achieve a top-5 rate of 14.8\% an improvement of the AlexNet. They were able to do this by tweaking the hyper-parameters of AlexNet while maintaining the same structure with additional deep learning elements. There is no record observed of use of ZFNet in agricultural systems despite the accuracy improvement. Each branch consists of three convolutional layers. The ‘first’ 1 x 1 projection reduces the dimensionality while the latter 1 x 1 projection expands the dimensionality. In between these convolutions, a regular (no annotation / asymmetric X), dilated (dilated X) or full convolution (no annotation) takes place. Batch normalization \cite{Ioffe_2015} and PReLU \cite{He_2015} are placed between all convolutions. As regularizer in the bottleneck, Spatial Dropout is used. MaxPooling on the master is added only when the bottleneck is downsampling which is true. 

GoogleNet, a 2014 ILSVRC image classification winner,  was inspired by LeNet but implemented a novel inception module. The Inception cell performs series of convolutions at different scales and subsequently aggregate the results. This module is based on several very small convolutions in order to drastically reduce the number of parameters. There has been tremedious efforts done to improve the performance of the architecture: a) Inception v1 \cite{Szegedy_2015} which performs convolution on an input, with 3 different sizes of filters (1x1, 3x3, 5x5). Additionally, max pooling is also performed. The outputs are concatenated and sent to the next inception module. b) Inception v2 and Inception v3 \cite{Szegedy_2016} factorize 5x5 convolution to two 3x3 convolution operations to improve computational speed. Although this may seem counterintuitive, a 5x5 convolution is 2.78 times more expensive than a 3x3 convolution. So stacking two 3x3 convolutions infact leads to a boost in performance. c) In Inception v4 and Inception-ResNet \cite{Szegedy_2017} the initial set of operations were modified before introducing the Inception blocks.  

Simonyan and Zisserman created VGGNet while doing investigation on the effect of convolutional network depth on its accuracy in the large-scale image recognition setting. The VGGNet took the second place after GoogLeNet in the competation. The model is made up of 16 convolutional layers which is similar to \cite{Alex_2012} but with many filters. There have been a number of update to the VGGNet archtecture starting with pioneer VGG-11(11 layers) which obtained 10.4\% error rate\cite{Simonyan_2014}. VGG-13 (13 layers) obtains 9.9\% error rate, which means the additional convolutional layers helps the classification accuracy. VGG-16(16 layers) obtained a 9.4\% error rate, which means the additional 3x1x1 conv layers help the classification accuracy. 1x1 convolution helps increase non-linearlity of the decision function, without changing the dimensions of input and output, 1x1 convolution is able to do the projection mapping in the same high dimensionality. This approach is used in NIN \cite{Lin_2013} GoogLeNet \cite{Szegedy_2015} and ResNet \cite{Kaiming_2015}. After updating to VGG-16 it obtained 8.8\% error rate which means the deep learning network was still improving by adding number of layers. VGG-19(19 layers) was developed to further improve the performance but it obtained 9.0\% showing no improvement even after adding more layers.  

When deeper networks starts converging, a degradation problem is exposed: with the network depth increasing, accuracy gets saturated and then degrades rapidly. Deep Residual Neural Network(ResNet) created by Kaiming He al. \cite{Kaiming_2015} introduced a norvel architecture with insert shortcut connections which turn the network into its counterpart residual version. This was a breakthrough which enabled the development of much deeper networks. The residual function is a refinement step in which the network learn how to adjust the input feature map for higher quality features. Following this intuition, the network residual block was refined and proposed a pre-activation variant of residual block \cite{He_2016}, in which the gradients can flow through the shortcut connections to any other earlier layer unimpeded. Each ResNet block is either 2 layer deep (used in small networks like ResNet 18, 34) or 3 layer deep(ResNet 50, 101, 152). This technique is able to train a network with 152 layers while still having lower complexity than VGGNet. It achieves a top-5 error rate of 3.57\% which beats human-level performance on this dataset. Although the original ResNet paper focused on creating a network architecture to enable deeper structures by alleviating the degradation problem, other researchers have since pointed out that increasing the network's width (channel depth) can be a more efficient way of expanding the overall capacity of the network.  

In DenseNet which is a logical extension of ResNet, there is improved efficiency by concatenating each layer feature map to every successive layer within a dense block \cite{Huang_2017}. This allows later layers within the network to directly leverage the features from earlier layers, encouraging feature reuse within the network. For each layer, the feature-maps of all preceding layers are used as inputs, and its own feature-maps are used as inputs into all subsequent layers, this helps alleviate the vanishing-gradient problem, feature reuse and reduce number of parameters.

\subsection{Proposed agricultural issues classification taxonomy} \label{subsec:taxonomy}

\begin{figure}[h!]
 \centering
  \includegraphics[height=14cm,height=7cm]{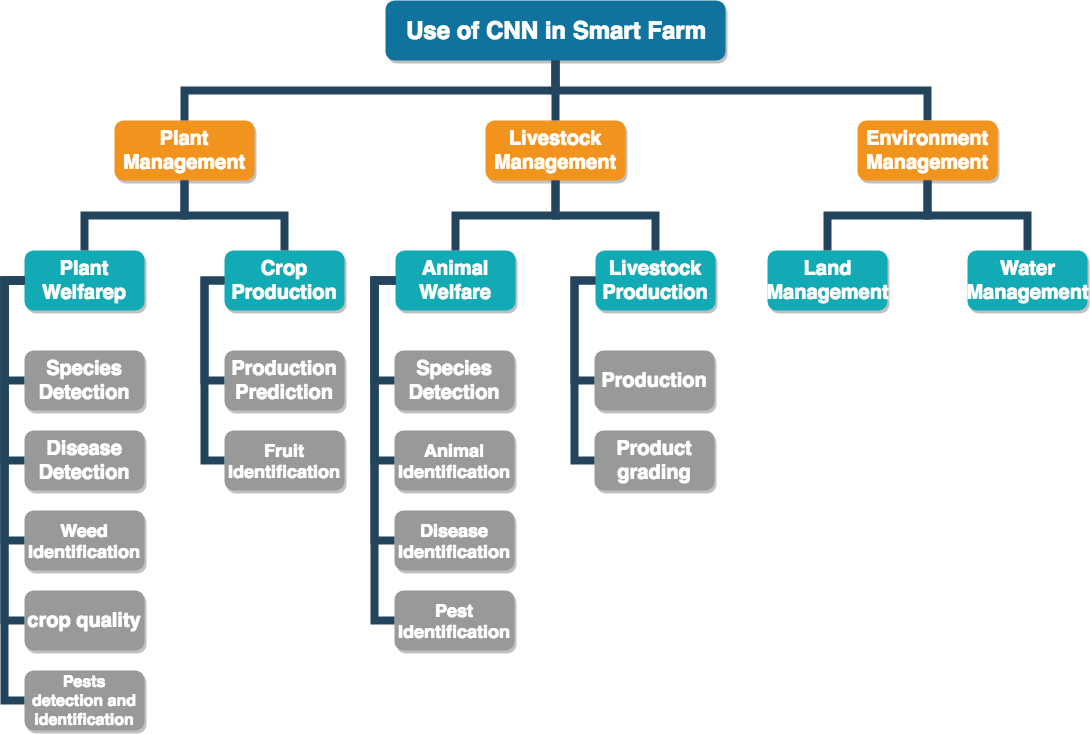}
 \caption{\small{Proposed classification taxonomy for CNN use in smart farm}}
 	\label{fig:taxonomy}
 \end{figure}

Many agricultural CNN solutions have been developed depending on specific agriculture issues. For the study purpose, a classification taxonomy tailored to CNN application in the smart farming was developed as shown Fig \ref{fig:taxonomy}. In this section, we categorize use of state-of-the-art CNN based on the agricultural issue they solve:

\begin{enumerate}[label=(\alph*)]
  \item Plant management includes solutions geared towards crop welfare and production. This includes classification(species), detection(disease and pest) and prediction(yield production).
  \item Livestock management address solutions for livestock production(prediction and quality management) and animal welfare(animal identification, species detection and disease and pest control).  
  \item Environment management addresses solutions for land and water management. 
\end{enumerate}

\subsection{Use of state-of-the-art CNN in Smart Farms} \label{subsec:cnnfarms} 
The table \ref{table:scnn}, shows use of state-of-the-art CNN in agriculture and in particular the areas of plant and leaf disease detection, animal face identification, plant recognition, land cover classification, fruit counting and identification of weeds. It consist of 5 columns to show: the problem description, size of data used, accuracy according to the metrics used, the state-of-the-art CNN used and reference literature. 

\begin{table}[h!]
\caption{\small{Use of state-of-the-art CNN in Smart Farm}}
\label{table:scnn}
\small
\begin{adjustbox}{width=1\textwidth}
\begin{tabular}
{|l|p{1.8in}|p{1in}|p{1.3in}|p{0.7in}|l|}

\hline
No. & Smartfarm Problem description                                          & Data used                                                                         & Accuracy                                                                           & CNN Framework used  & Article                                    \\ \hline
1   & Fruit detection                                                        & Images of three  fruit varieties: apples (726), almonds (385) and mangoes  (1154) & F1 (precicion score) of 0.904 (apples) 0.908 (mango) 0.775 (almonds)                          & VGGNet              & \cite{Bargoti_2017}      \\ \hline
2   & Detection of sweet pepper  and rock melon fruits                       & 122 images                                                                        & 0.838 (F1)                                                                         & VGGNet              & \cite{Sa_2017}           \\ \hline
3   & Recognize different plant species                                      & Data set of 44 classes                                                            & 99.60\% (CA - correct prediction)                                                                       & AlexNet             & \cite{Lee_2015}          \\ \hline
4   & Recognize different plant                                              & 91 759 images                                                                     & 48.60\% (LC-correct species classification)                                                                       & AlexNet             & \cite{Reyes_2015}        \\ \hline
5   & Identify obstacles  in row crops and grass mowing                      & 437 images                                                                        & 99.9\% in row crops and 90.8\% in grass mowing (CA)                                & AlexNet             & \cite{Steen_2016}        \\ \hline
6   & Identify crop species and diseases                                     & 54 306 images                                                                     & 0.9935 (F1)                                                                        & AlexNet + GoogLeNet & \cite{Mohanty_2016}      \\ \hline
7   & Detect obstacles  that are distant,  heavily occluded  and unknown     & 48 images                                                                         & 0.72 (F1)                                                                          & AlexNet + VGG       & \cite{Christiansen_2016} \\ \hline
8   & Leaf disease detection                                                 & 4483 images                                                                       & 96.30\% (CA)                                                                       & CaffeNet            & \cite{Sladojevic_2016}   \\ \hline
9   & Identify thistle in winter wheat and spring barley images              & 4500 images                                                                       & 97.00\% (CA)                                                                       & DenseNet            & \cite{Sorensen_2017}     \\ \hline
10  & Predict number of tomatoes in images                                   & 24 000 images                                                                     & 91\% (RFC-Ratio of total fruits counted) on real images, 93\% (RFC) on synthetic  images & GoogLeNet + ResNet  &  \cite{Rahnemoonfar_2017} \\ \hline
11  & Classify banana leaf diseases                                          & 3700 images                                                                       & 96\% (CA), 0.968 (F1)                                                              & LeNet               &  \cite{Amara_2017}        \\ \hline
12  & Indentify pig face                                                     & 1553 images                                                                       & 96.7\%( CA)                                                                        & VGGNet              &  \cite{Hansen_2018}       \\ \hline
13  & Classify weed from crop species based on 22 different species in total & 10413 images                                                                      & 86.20\% (CA)                                                                       & VGGNet              &  \cite{Dyrmann_2016}      \\ \hline

14  & Detecting and categorizing the criticalness of Fusarium wilt of radish based on thresholding a range of color features & 1500 images                                                                      &  & GoogLeNet              &  \cite{Hyun_2018}      \\ \hline

15  & Fruit counting & 24 000 images  & 91\% accuracy & Inception-ResNet              &  \cite{Rahnemoonfar_2017} \\ \hline

16  & Automatic Plant disease diagnosis for early disease symptoms & 8178 images & overall improvement of the balanced accuracy from 0.78 to 0.87 from previous 2017 study & Deep ResNet              &  \cite{Picona_2018} \\ \hline

\end{tabular}

\end{adjustbox}

\end{table}

In their paper, Amara et al. \cite{Amara_2017} use the LeNet architecture to classify the banana leaves diseases. The model was able to effectively classify the leaves after several experiments. The approach was able to classify leaves images with different illumination, complex background, resolution, size, pose, and orientation. We also reviwed use of CaffeNet archtecture \cite{Jia_2014} in agricultural application,  which is a 1-GPU version of AlexNet. The success of this model at LSVRC 2012 \cite{LSVRC_2012} encourage many computer vision community to explore more on the application of deep learning in computer vision. Mohanty et al. \cite{Mohanty_2016} combined both AlexNet and GoogLeNet to identify 14 crop species and 26 diseases(or absence thereof) from a dataset of 54,305 images. The approach records an impressive accuracy of 99.35\% demonstrating the feasibility of the state-of-the-art CNN architectures. Other areas AlexNet has been used with high accuracy record include; identify plants using different plant views \cite{Reyes_2015}, identify plant species \cite{Lee_2015}, identify obstacles in the farm \cite{Steen_2016} and leaf disease detection \cite{Sladojevic_2016}. Because of its achievement to improve utilization of the computing resources GoogLeNet has been used  in fruit count \cite{Rahnemoonfar_2017} and plant species classification \cite{Mohanty_2016}. VGGNet has been used in classifying weed \cite{Dyrmann_2016}, detect obstacles in the farm \cite{Christiansen_2016}, fruit detection \cite{Sa_2017} and animal face recognition \cite{Hansen_2018}. Like ResNet, DenseNet is a recent model that explains why it has not been employeed significantly in farming, nevertheless it has been used in thistle identification in winter wheat and spring barley \cite{Sorensen_2017}. Since ResNet is a such a recent model, it have only been used by one author in fruit counting \cite{Rahnemoonfar_2017}.
Many of the CNN developed for agricultural use depend on the problem or challenge they solve.  

\section{Conclusions and Recommendations} \label{sec:discussions}
Despite remarkable achievement in use state-of-the-art CNN in agriculture in general, there exist grey areas in relation to smart farm that future researchers may look at. These areas may include; real-time image classification, interactive image classification and interactive object detection. State-of-the-art CNN is relatively a new technology that explain why the finding of the study about their use in smart farm is relatively small. However, its is important to note that models built from state-of-the-art architectures have a impressive record of better precision performance. In this paper, we aimed at establishing the potential of state-of-the-art CNN in IoT based smart farms. In particular we first discussed the architectures of state-of-the-art CNNs and their respective prediction accuracy at the ILSVRC challenge. Then a survey on application of the identified CNNs in Agriculture was performed; to examine the particular application in a smart farm, listed technical details of the architecture employed and overall prediction accuracy achieved according to the author precision metrics.  
From the study its evident of continuous accuracy improvement of the state-of-the-art CNN architectures as computer vision community put effort to perfect the methods. The findings indicate that state-of-the-art CNN has achieved better precision in all the cases applied in the agricultural domain, scoring higher accuracy in majority of the problem  as compared to other image-processing techniques. Considering that the state-of-the-art CNN has achieved state-of-the-art results in prediction in general and high precision in the few farming cases observed, there is great potential that can be achieved in using the methods in smart farming. It has been observed that many authors apply more than one architecture in order to optimize the performance of the network without compromising the expected accuracy. This approach is very efficient in the observed cases, and we recommend similar hybrid approach when building robust IoT based networks which are computationally fair to the mobile devices. This study aims to motivate researchers to experiment and apply the state-of-the-art methods in smart farms problems related to computer vision and data analysis in general.

%
%

\end{document}